\RequirePackage{lineno}

\documentclass[prd,twocolumn,showpacs,superscriptaddress,amsmath,amssymb,nofootinbib]{revtex4-1}
\usepackage{graphicx}
\usepackage{epsfig}
\usepackage{overpic}
\usepackage{dcolumn}
\usepackage{bm}
\usepackage{color}
\usepackage{lineno}
\usepackage{xspace}
\usepackage{ctable}
\usepackage{verbatim}
\usepackage{subfigure}
\include{def-com}
\usepackage{hyperref}
\hypersetup{colorlinks,linkcolor={blue},citecolor={blue},urlcolor={red}}

\usepackage{booktabs,tabularx}
\usepackage{array}
\newcolumntype{L}[1]{>{\raggedright\let\newline\\\arraybackslash\hspace{0pt}}m{#1}}
\newcolumntype{C}[1]{>{\centering\let\newline\\\arraybackslash\hspace{0pt}}m{#1}}
\newcolumntype{R}[1]{>{\raggedleft\let\newline\\\arraybackslash\hspace{0pt}}m{#1}}
\usepackage{titlesec}

\newcommand{\mevcc}{\ensuremath{\;\text{MeV}/c^2\xspace}}

\newcommand{\gev}{\ensuremath{\;\text{GeV}\xspace}}

\begin{document}
\title{\boldmath First  Measurements of $\chi_{cJ}\rightarrow \Sigma^{-} \bar{\Sigma}^{+} (J = 0, 1, 2)$  Decays }

\author{
  \begin{small}
    \begin{center}
M.~Ablikim$^{1}$, M.~N.~Achasov$^{10,d}$, P.~Adlarson$^{64}$, S. ~Ahmed$^{15}$, M.~Albrecht$^{4}$, A.~Amoroso$^{63A,63C}$, Q.~An$^{60,48}$, ~Anita$^{21}$, Y.~Bai$^{47}$, O.~Bakina$^{29}$
% ...
\\
\vspace{0.2cm}
(BESIII Collaboration)\\
\vspace{0.2cm} {\it
$^{1}$ Institute of High Energy Physics, Beijing 100049, People's Republic of China\\
% ...
$^{4}$ Bochum Ruhr-University, D-44780 Bochum, Germany\\
% ...
$^{10}$ G.I. Budker Institute of Nuclear Physics SB RAS (BINP), Novosibirsk 630090, Russia\\
% ...
$^{15}$ Helmholtz Institute Mainz, Johann-Joachim-Becher-Weg 45, D-55099 Mainz, Germany\\
% ...
$^{21}$ Indian Institute of Technology Madras, Chennai 600036, India\\
% ...
$^{29}$ Joint Institute for Nuclear Research, 141980 Dubna, Moscow region, Russia\\
% ...
$^{47}$ Southeast University, Nanjing 211100, People's Republic of China\\
$^{48}$ State Key Laboratory of Particle Detection and Electronics, Beijing 100049, Hefei 230026, People's Republic of China\\
% ...
$^{60}$ University of Science and Technology of China, Hefei 230026, People's Republic of China\\
% ...
$^{63}$ (A)University of Turin, I-10125, Turin, Italy; (B)University of Eastern Piedmont, I-15121, Alessandria, Italy; (C)INFN, I-10125, Turin, Italy\\
$^{64}$ Uppsala University, Box 516, SE-75120 Uppsala, Sweden\\
% ...
$^{d}$ Also at the Novosibirsk State University, Novosibirsk, 630090, Russia\\
% ...
}\end{center}
\vspace{0.4cm}
\end{small}
}
\noaffiliation{}

\begin{abstract}
We measured the branching fractions of the decays $\chi_{cJ}\to\Sigma^{-}\bar{\Sigma}^{+}$  for the first time using the final states $n\bar{n}\pi^{+}\pi^{-}$. The data sample  exploited here is $448.1\times10^{6}$ $\psi(3686)$ events collected with BESIII. We find $\mathcal{B}(\chi_{cJ}\rightarrow\Sigma^{-}\bar{\Sigma}^{+}) = (51.3\pm2.4\pm4.1)\times10^{-5},\, (5.7\pm1.4\pm0.6)\times10^{-5},\, \rm{and}~ (4.4\pm1.7\pm0.5)\times10^{-5}$, for $J=0,1,2$, respectively, where the first uncertainties are statistical and the second systematic.
  
\end{abstract}

\maketitle

\section{\boldmath INTRODUCTION}

Experimental studies of the $\chi_{cJ}(J=0,1,2)$ states are important for testing models that are based on
non-perturbative Quantum Chromodynamics~(QCD). The $\chi_{cJ}$ mesons are $P$-wave $c\bar{c}$ triple-states with a
spin parity $J^{++}$, and cannot be produced directly in $e^+e^-$ annihilation. However, they can be produced
in the radiative decays of the vector charmonium state $\psi(3686)$ with considerable branching fractions (BFs) of $\sim9\%$~\citep{PDG}.
A large sample of $\psi(3686)$ decays has been collected at BESIII, which provides a good opportunity to investigate
the $P$-wave $\chi_{cJ}$ states~\cite{pwave}.

Many theoretical calculations show that the color octet mechanism (COM) could have a large contribution in describing
$P$-wave quarkonium decays~\citep{COM,ppbar,COM2}. The predictions for $\chi_{cJ}$ decays to meson pairs are in agreement
with the experimental results~\citep{meson}, while contradictions are observed in the $\chi_{cJ}$ decays to baryon pairs
$(B\bar{B})$~\citep{ppbar,COM2}. For example, the predicted BFs of $\chi_{cJ}\to\Lambda\bar{\Lambda}$ disagree with measured values~\citep{BES}.
In addition, the study of $\chi_{c0}\to B \bar{B}$ is helpful to test the validity of the helicity selection rule (HSR)~\citep{HSR,HSR1},
which prohibits $\chi_{c0}\to B\bar{B}$. Measured BFs for $\chi_{c0}\to p\bar{p},\,\Lambda\bar{\Lambda}~\rm{and}~\Xi^{-}\bar{\Xi}^{+}$
do not vanish~\citep{BES,BES_hsr1}, demonstrating a strong violation of HSR in charmonium decay.
The quark creation model (QCM)~\citep{QCM} is developed to explain the strengthened decays of $\chi_{c0}\to B \bar{B}$ and it
predicts the rate of $\chi_{c0,2 }\to \Xi^+\Xi^-$~\citep{BES_hsr1} well. However, the same model is unable to accurately reproduce
the observed decay rates to other $B\bar{B}$ final states~\citep{BES}. Recent BF data for $\chi_{c1,2}\to\Sigma^{+}\bar{\Sigma}^{-}$ and $\Sigma^{0} \bar{\Sigma}^{0}$~\citep{CLEO} are in good agreement with COM predictions~\citep{ppbar}, while measured BFs of $\chi_{c0}\to\Sigma^{+}\bar{\Sigma}^{-}$ and $\Sigma^{0} \bar{\Sigma}^{0}$~\citep{CLEO, BES2} are inconsistent with COM models based on the charm-meson-loop mechanism~\citep{COM2,Charm}, and violate the HSR, too. Experimentally, there are no BF data of
$\chi_{cJ}\to\Sigma^{-}\bar{\Sigma}^{+}$, and therefore those measurements are necessary to further test the validity of COM, HSR and QCM. 

In this paper, we report on an analysis of the processes $\psi(3686)\rightarrow\gamma\chi_{cJ}$, $\chi_{cJ}\rightarrow\Sigma^{-}\bar{\Sigma}^{+}$ ($\Sigma^{-}\to n\pi^{-}$, $\bar{\Sigma}^{+}\to\bar{n}\pi^{+}$) using a data sample of $(448.1\pm 2.9)\times10^{6}$ $\psi(3686)$ events collected with BESIII~\citep{psidata}. The BFs of the decays $\chi_{cJ}\to\Sigma^{-}\bar{\Sigma}^{+}$  are measured for the first time. 

\section{\boldmath BESIII DETECTOR AND MONTE CARLO SIMULATION}

The BESIII detector operating at the Beijing electron-positron collider (BEPCII), is a double-ring $e^{+}e^{-}$ collider with a peak luminosity
of $1 \times10^{33}\textrm{\ cm}^{-2}\textrm{s}^{-1}$ at center-of-mass energy $\sqrt{s}=3.77~\mathrm{GeV}$~\citep{pwave, BESDetector}.
The BESIII detector has a geometrical acceptance of $93\%$ over $4\pi$ solid angle. The cylindrical core of the BESIII detector consists
of a small-cell, helium-gas-based ($60\%$ He, $40\%$ C$_{3}$H$_{8}$) main drift chamber (MDC) which is used to track the charged particles.
The MDC is surrounded by a time-of-flight (TOF) system built from plastic scintillators that is used for charged-particle identification (PID).
Photons are detected and their energies and positions are measured with an electromagnetic calorimeter (EMC) consisting of 6240 CsI(TI) crystals.
The sub-detectors are enclosed in a superconducting solenoid magnet with a field strength of 1~T. Outside the magnet coil, the muon detector
consists of 1000~m$^{2}$ resistive plate chambers in nine barrel and eight end-cap layers, providing a spatial resolution of better than 2~cm.
The momentum resolution of charged particle is $0.5\%$ at 1\gev. The energy loss ($dE/dx$) measurement provided by the MDC has a resolution of 6\%,
and the time resolution of the TOF is 80~ps (110~ps) in the barrel (end-caps). The energy resolution for photons is 2.5\% (5\%) at 1~GeV in the barrel (end-caps) of the EMC. 

A dedicated Monte Carlo (MC) simulation of the BESIII detector based on \textsc{geant4}~\citep{geant4}
is used for the optimization of event selection criteria, the determination of the detection efficiencies,
and to estimate the contributions of backgrounds. A generic MC sample with~$5.06\times10^{8}$ events is generated,
where the production of the $\psi(3686)$ resonance is simulated by the MC event generator \textsc{ kkmc}~\citep{KKMC}.
Particle decays are generated by \textsc {evtgen}~\citep{BESEVTGEN} for the known decay modes with BFs taken from Particle Data Group (PDG),
and by \textsc{lundcharm}~\citep{LundCharm} for the remaining unknown decays.
For the MC simulation of the signal process, the decay of $\psi(3686)\rightarrow\gamma\chi_{cJ}$ is generated by following the angular
distributions taken from Ref.~\citep{Angular}, where the polar angles $\theta$ of radiation photons are distributed according to
$(1+\cos^{2}\theta), (1-\frac{1}{3}\cos^{2}\theta),(1+ \frac{1}{13} \cos^{2}\theta)$ for $J = 0,~1,~2$.
The $\chi_{cJ}\to\Sigma^{-}\bar{\Sigma}^{+}$ decays are generated with the \textsc{angsam}~\citep{BESEVTGEN} model,
with helicity angles of the $\Sigma$ satisfying the angular distribution $1+\alpha\cos^{2}\theta$. Note that $\alpha=0$ for the decay of the $\chi_{c0}$ because the helicity angular distribution of a scalar particle is isotropic. The subsequent decays $\Sigma^{-}\to n\pi^{-}$ and $\bar{\Sigma}^{+}\to \bar{n}\pi^{+}$ are generated with uniform momentum distribution in the phase space (\textsc{PHSP})~\citep{explan}.

\section{\boldmath EVENT SELECTION AND BACKGROUND ANALYSIS}

We reconstruct the candidate events from the decay chain $\psi(3686)\to\gamma\chi_{cJ}$ followed by
$\chi_{cJ}\to\Sigma^{-}\bar{\Sigma}^{+}$ with subsequent decays $\Sigma^{-}\to n\pi^{-}$ and $\bar{\Sigma}^{+}\to \bar{n}\pi^{+}$.
The charged tracks are reconstructed with the hit information from the MDC. The polar angles of charged tracks in the MDC have to
fulfill $|\cos\theta|<0.93$. A loose vertex requirement is applied for charged-track candidates to implement the sizable decay lengths of
$\Sigma^{-}~\rm{and}~\bar{\Sigma}^{+}$, and each charged track is required to have a point of closest approach to $e^{+}e^{-}$ interaction point that is within 10~cm in the plane perpendicular to the beam axis and within $\pm 30$~cm in the beam direction. The combined information of $dE/dx$ and TOF is used to calculate PID probabilities for the pion, kaon and proton hypothesis, respectively, and the particle type with the highest probability is assigned to the corresponding track. In this analysis, candidate events are required to have two charged tracks identified as $\pi^{+}$ and $\pi^{-}$.

There are three neutral particles in the final states of the signal process, the radiative photon $\gamma$, anti-neutron $\bar{n}$ and neutron $n$.
The radiative photon deposits most of its energy in the EMC with a high efficiency. The $\bar{n}$ annihilates in the EMC and produces several
secondary particles with a total energy deposition up to 2~GeV. The $n$, on the other hand, is not identifiable due to its low interaction
efficiency and its small energy deposition. Therefore, the $\bar{n}$ and radiative photon are selected in this process. The most energetic shower in the EMC is assigned to be the $\bar{n}$ candidate. To discriminate $\bar{n}$ from photons and to suppress the electronic noise, several selection criteria are used. Firstly, the deposited energy of $\bar{n}$ is required to be in the range 0.2$-$2.0~GeV. 
Secondly, the second moment of candidate shower, defined as $S=\sum_{i}E_{i}r_{i}^{2}/\sum_{i}E_{i}$, 
must satisfy $S>20$~cm$^{2}$, where $E_{i}$ is the energy deposited in the $i^{th}$ crystal of the shower and $r_{i}$ is 
the distance from the center of that crystal to the center of the shower~\citep{nnbar}. Furthermore, the number of EMC hits in a 40$^{\circ}$ cone seen from the vertex around the $\bar{n}$ shower direction is required to be greater than $20$. After applying these selection criteria, the $\bar{n}$ candidates have a purity of more than 98\% estimated from signal MC sample.

To avoid the secondary showers originating from $\bar{n}$ annihilation, the radiative photon is selected from EMC showers that have an
opening angle with respect to the $\bar{n}$ direction that is greater than $40^{\circ}$. Good photon candidates are selected by requiring a minimum energy deposition of 80~MeV in the EMC, and are isolated from all charged tracks by a minimum angle of $10^{\circ}$. The time information of the EMC is used to further suppress electronic noise and energy depositions unrelated to the event. At least one good photon candidate is required in an event.

The momentum or direction information of candidate particles are subjected to a kinematic fit that assumes the $\psi(3686)\to\gamma n\bar{n}\pi^{+}\pi^{-}$ hypothesis, where the direction of $\bar{n}$ in the fit is involved and $n$ is treated as a missing particle. The kinematic fit is then applied by
imposing energy and momentum conservation at the IP and by constraining the $\bar{n}\pi^{+}$ invariant mass to match the nominal $\bar{\Sigma}^{+}$ mass~\cite{PDG}. For events with more than one photon candidate, the combination with a minimum $\chi^{2}_{\rm kfit}$ is chosen with the requirement that $\chi^{2}_{\rm kfit}<20$.

\begin{figure*}[hbtp]
\centering
\begin{overpic}[width=8.7cm,angle=0]{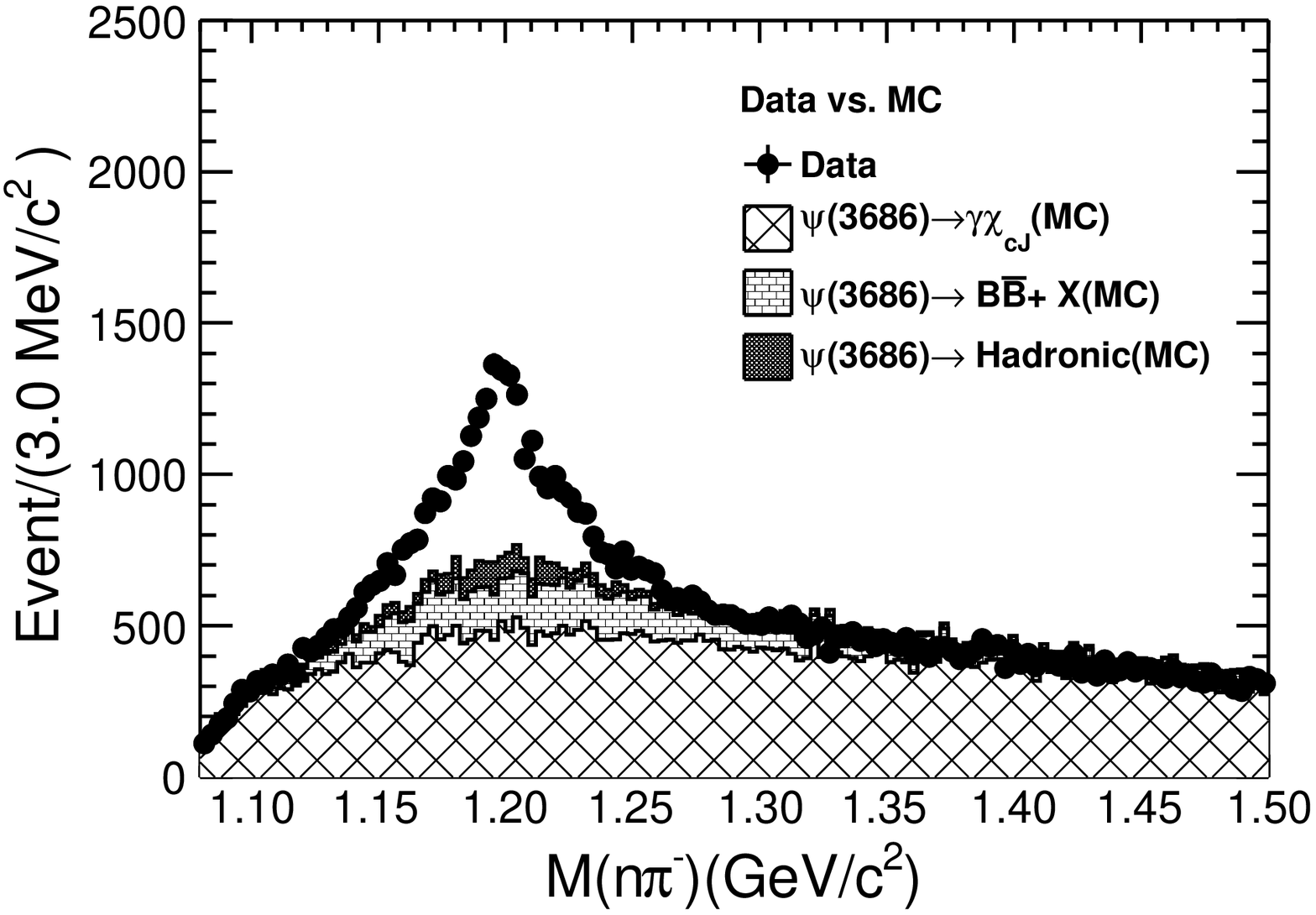}
\put(85,55){\textbf{(a)}}
\end{overpic}
\begin{overpic}[width=8.7cm,angle=0]{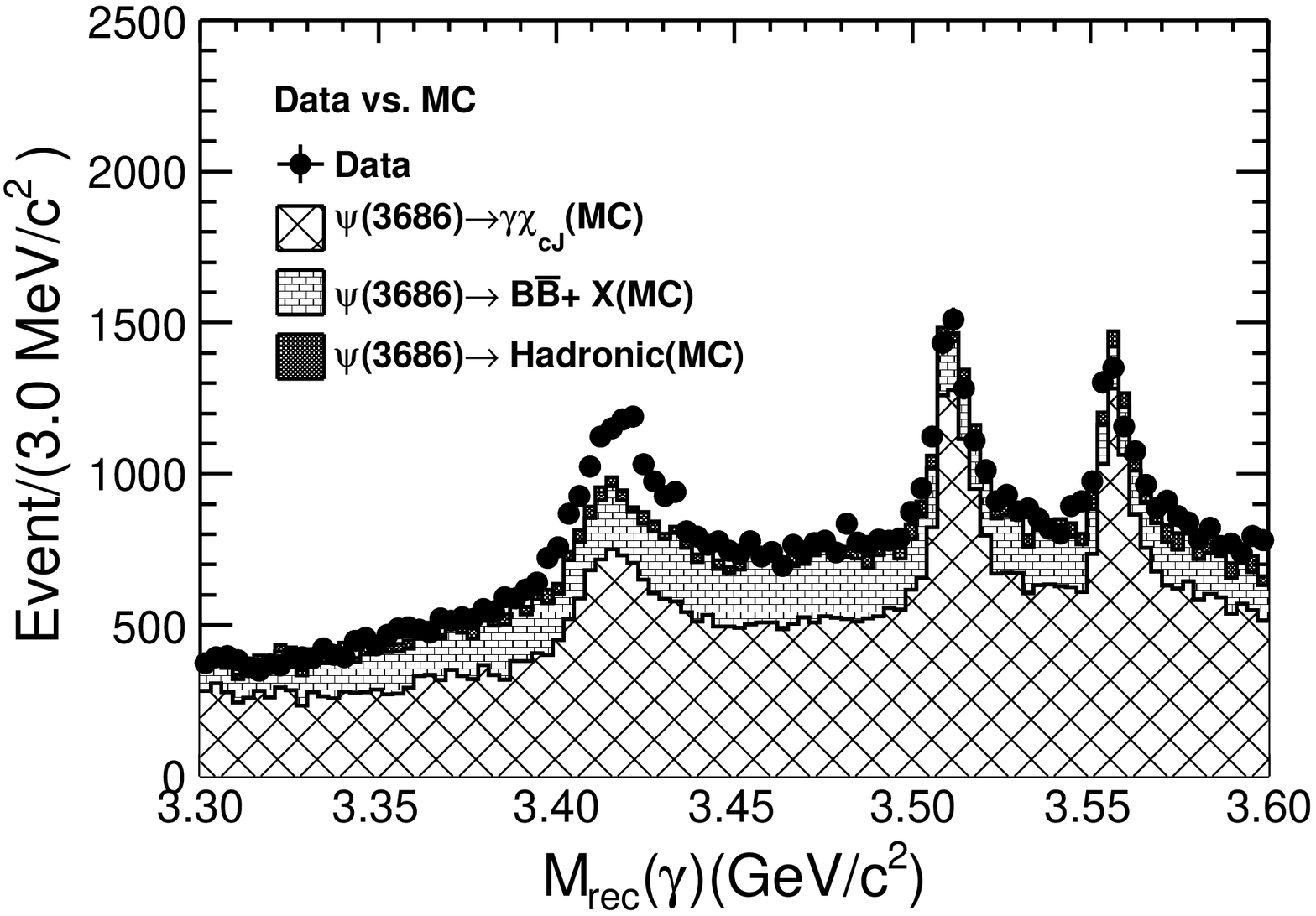}
\put(85,55){\textbf{(b)}}
\end{overpic}
\vspace*{-0.3cm}
\caption{Invariant-mass distributions of reconstructed $\Sigma^{-}$ candidates (a) and the recoil mass of $\gamma$ (b).
  The dots with error bars denote the data. The contributions for each component are obtained using MC simulations and are indicated
  as the hatched histograms.}
\label{invmass}
\end{figure*}

After applying the kinematic fit, the backgrounds from $\psi(3686)\to\pi^{0}\pi^{0}J/\psi$ followed by decays of $J/\psi\to B\bar{B}$ and $\pi^{0}\to\gamma\gamma$ are suppressed by reconstructing events with two photon candidates.
An event is then discarded when the invariant mass of any two photons are located within 120~MeV/$c^{2}$ and 150~MeV/$c^{2}$.
The contamination of the channel $\psi(3686)\to\pi^{+}\pi^{-}J/\psi$ with $J/\psi\to n\bar{n}$ is removed by requiring
$|M_{\rm rec}(\pi^{+}\pi^{-})-m(J/\psi)|>10$~MeV/$c^{2}$, where $M_{\rm rec}(\pi^{+}\pi^{-})$ is the recoil mass of the $\pi^{+}\pi^{-}$ pair and $m(J/\psi)$ is the world average mass of the $J/\psi$ meson~\cite{PDG}. Another sources  of backgrounds are from events containing a $K_{\rm S}^{0}$. These events are removed by requiring $|M(\pi^{+}\pi^{-})-m(K_{\rm S}^{0})|>10~$MeV/$c^{2}$, whereby $M(\pi^{+}\pi^{-})$ and $m(K_{\rm S}^{0})$ are the reconstructed $\pi^+\pi^-$ invariant mass and world average mass of the $K_{\rm S}^{0}$~\cite{PDG}, respectively.
The signal could be contaminated with background from $\psi(3686)\to\Sigma^{-}\bar{\Sigma}^{+}$ whereby one fake photon has been
reconstructed. To remove such background, events are rejected for which the $\chi^{2}_{\rm kfit}(\Sigma^{-}\bar{\Sigma}^{+})$ is smaller than
$\chi^{2}_{\rm kfit}(\gamma\Sigma^{-}\bar{\Sigma}^{+})$.

\begin{figure}[htbp]
\begin{center}
\begin{overpic}[width=8.8cm,angle=0]{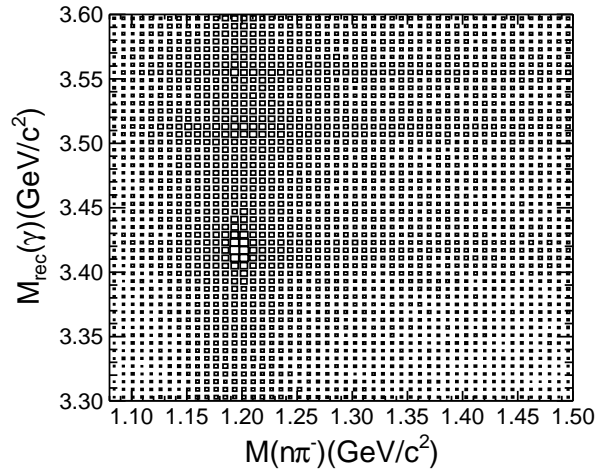}
\end{overpic}
\end{center}
\vspace*{-0.5cm}
\caption{A 2D distribution of $M_{\rm rec}(\gamma)$ versus $M(n\pi^{-})$ for data.}
\label{2Dscatter}
\end{figure}

The invariant-mass spectrum of $n\pi^{-}$ and the recoil mass spectrum of the $\gamma$ are shown in Fig.~\ref{invmass} for both data and  MC simulations, where $\Sigma^{-}$ and $\chi_{cJ}$ signals can be observed. The MC results represent the main characteristics of the various background sources. However, they cannot fully describe the data due to missing or improper modeling of background processes involving $B\bar{B}$, especially when the final states contain $n\bar{n}$. Using the topology technique~\citep{gemc}, we have categorized the main background sources into three kinds:
a) the process $\psi(3686)\to\gamma\chi_{cJ}$ whereby the $\chi_{cJ}$ decays to hadronic final states, which shows a peak in
$M_{\rm rec}(\gamma)$ and no peaking structure in $M(n\pi^{-})$;
b) the process $\psi(3686)\to B\bar{B}$ or $J/\psi\to B\bar{B}$ via the hadronic transition from $\psi(3686)$, which is not peaking in
$M_{\rm rec}(\gamma)$ but shows a wide bump in $M(n\pi^{-})$;
c) the decays $\psi(3686)$ to hadronic final states, which are non-peaking in both $M_{\rm rec}(\gamma)$ and $M(n\pi^{-})$.
Besides, a two-dimensional (2D) distribution of $M(n\pi^{-})$ and $M_{\rm rec}(\gamma)$ is shown in Fig.~\ref{2Dscatter} for the data.
Clear accumulations of candidate events of the signal process $\chi_{c0}\to \Sigma^{-}\bar{\Sigma}^{+}$ are observed around the
intersections of the $\chi_{c0}$ and $\Sigma^{-}$ mass regions, and a signature of the process $\chi_{c1,2}\to\Sigma^{-}\bar{\Sigma}^{+}$
can be observed. A data sample corresponding to an integrated luminosity of 44~pb$^{-1}$, taken at $\sqrt{s}=3.65$~GeV, is used to estimate the continuum background arising from quantum electrodynamics (QED) processes. No peaking backgrounds are observed in the mass spectrum of $M_{\rm rec}(\gamma)$ for the continuum data sample, therefore the contribution from QED background can be neglected.

\section{\boldmath EXTRACTION OF THE SIGNAL}
\label{extraction}
To extract the signal yields for $\chi_{cJ}\to\Sigma^{-}\bar{\Sigma}^{+}$, unbinned maximum-likelihood fits to the $M_{\rm rec}(\gamma)$
distributions as a function of $M(n\pi^{-})$ are performed, noted as bin-by-bin fit. The bin width for $M(n\pi^{-})$ is determined by testing the MC samples,
where the MC samples include events from MC-generated background sources, and events randomly sampled from signal MC events with the same amount events as observed
in data as signal. The bin width is determined when the minimum input-output difference is obtained for the extraction of the signal and it is found to be 10~MeV/$c^{2}$.

\begin{figure}[htbp]
\begin{center}
\begin{overpic}[width=8.7cm,angle=0]{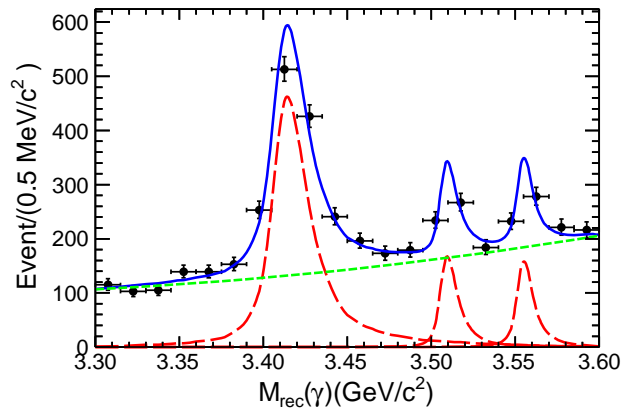}
\end{overpic}
\end{center}
\vspace*{-0.5cm}
\caption{Fit to the $M_{\rm rec}(\gamma)$ distribution at the maximum accumulation in the $M(n\pi^{-})$ bin. Black dots with error bars are from data, the solid blue
  lines are the best fit result, dashed red lines represent signal contributions, and dashed green lines are the fitted backgrounds.}
\label{fitchicj}
\end{figure}

\begin{figure*}[htbp]
\begin{center}
\begin{overpic}[width=5.9cm,angle=0]{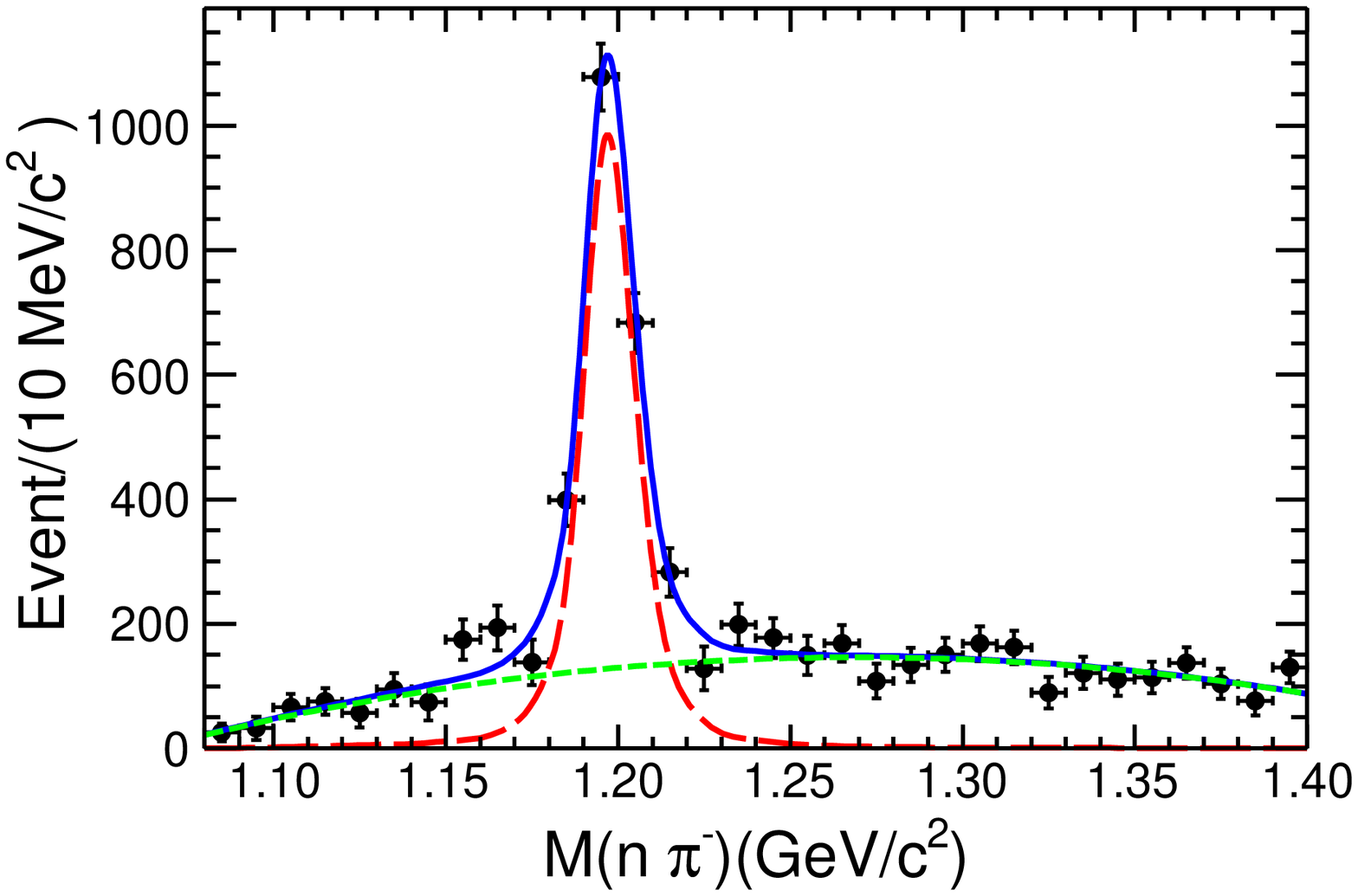}
\put(80,55){\textbf{(a)}}
\end{overpic}
\begin{overpic}[width=5.9cm,angle=0]{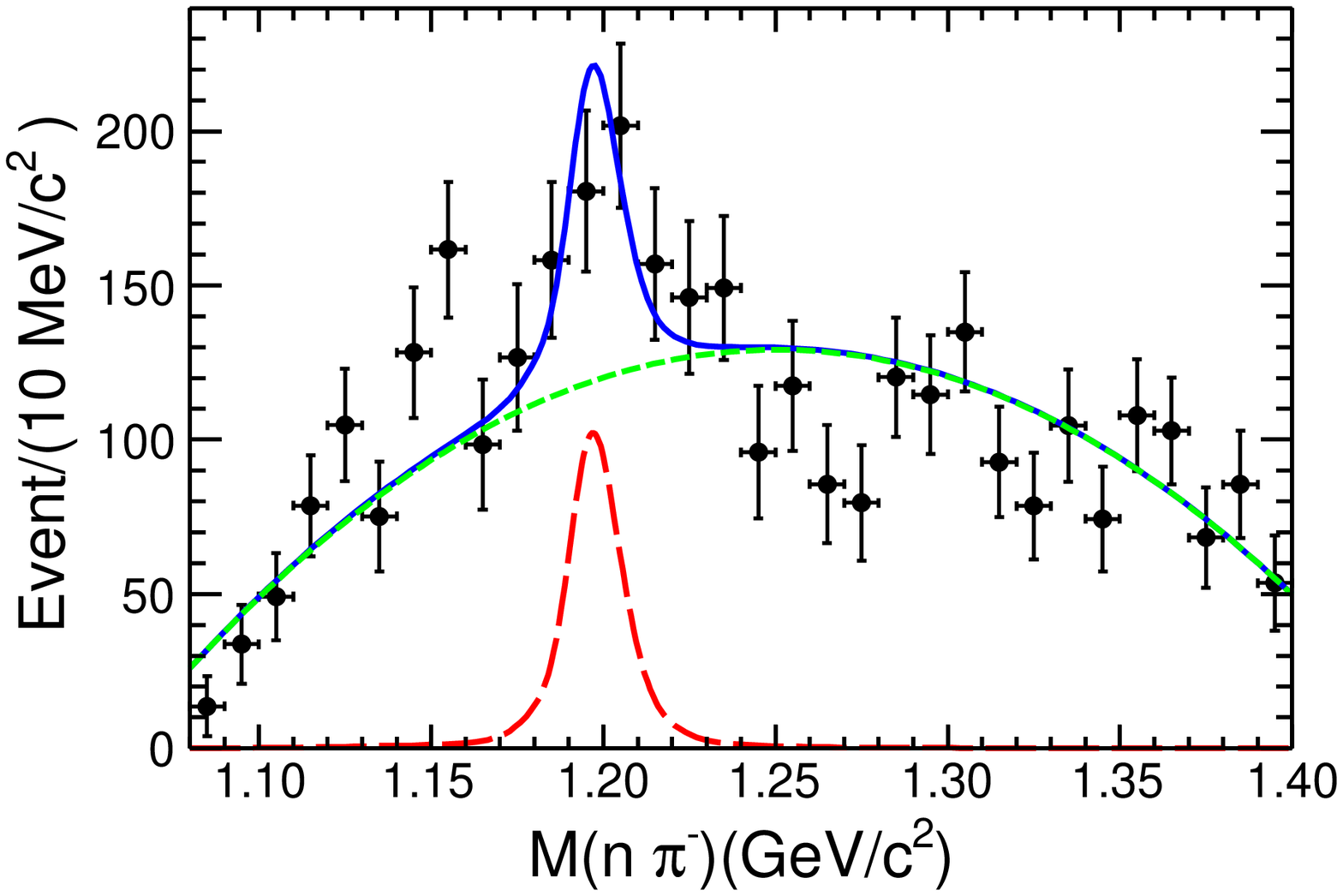}
\put(80,55){\textbf{(b)}}
\end{overpic}
\vspace{0.065cm}
\begin{overpic}[width=5.9cm,angle=0]{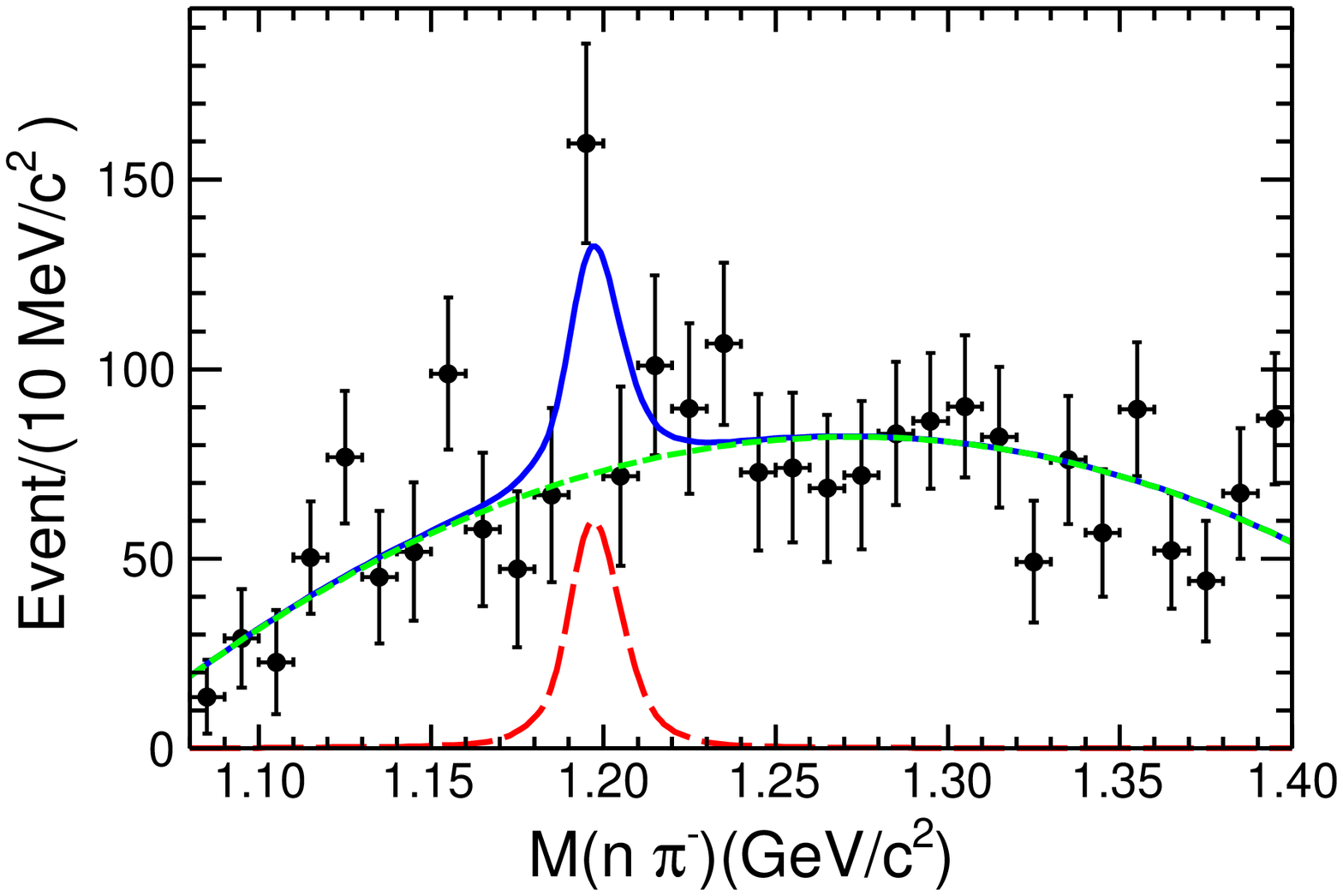}
\put(80,55){\textbf{(c)}}
\end{overpic}
\end{center}
\vspace*{-0.5cm}
\caption{The $\chi_{cJ}\to\Sigma^{-}\bar{\Sigma}^{+}$ signal yields as a function of $M(n\pi^{-})$ for (a) $\chi_{c0}$ (b) $\chi_{c1}$ and (c) $\chi_{c2}$.
  Black dots with error bars correspond to data, the solid blue lines are the overall fit results, dashed red lines represent signal contributions,
  and dashed green lines are the fitted backgrounds.}
	\label{fit}
\end{figure*}

In the fit of $M_{\rm rec}(\gamma)$ in each $n\pi^{-}$ bin, the $\chi_{cJ}$ signals are described by the MC shapes convoluted with Gaussian functions
to compensate for a possible resolution difference between the data and MC.
For a proper modeling of the lineshape of the signal, thereby suppressing photon misidentification, we selected signal MC events for which the opening angle of the reconstructed photon matches the value given by the generator. A second-order Chebychev polynomial function is used to describe the non-$\chi_{cJ}$ background. It should be noted that the $M_{\rm rec}(\gamma)$ resolution of the process $\psi(3686)\to\gamma\chi_{cJ}$, with inclusive decays of the $\chi_{cJ}$, is the same as observed in the signal MC data. Figure~\ref{fitchicj} shows the results of a bin-by-bin fit of one of the $M_{\rm rec}(\gamma)$ distributions selected for a bin in $M(n\pi^{-})$ at the $\Sigma^-$ peak position. Figure~\ref{fit} shows the fitted signal yields of $\psi(3686)\to\gamma \chi_{cJ}$ as a function of $M(n\pi^{-})$. Clear signatures of $\Sigma^-$ decays can be observed. Binned least-$\chi^{2}$ fits are subsequently performed to these spectra. The signal shapes are described by MC-simulated responses convoluted with Gaussian distributions and backgrounds are described by second-order Chebychev polynomials. The fit results are shown by the lines in Fig.~\ref{fit}. The statistical significances of the signal for the three $\chi_{cJ}$ cases are found to be 30$\sigma$, 5.8$\sigma$ and 3.6$\sigma$, respectively. The significances are calculated from the $\chi^{2}$ differences between fits with and without the signal processes. The corresponding signal yields are summarized in Table~\ref{sum}.
The BFs are obtained from:

\begin{equation}
\mathcal{B}(\chi_{cJ}\rightarrow \Sigma^{-}\bar{\Sigma}^{+}) =\frac{N^{obs}}{N_{\psi(3686)}\cdot\epsilon\cdot \prod\mathcal{B}_{i}}~,
\end{equation}
where $N^{obs}$ is the number of signal events obtained from the bin-by-bin fit; $\epsilon$ is the detection efficiency obtained from signal MC after the photon matching;
$\prod\mathcal{B}_{i}$ is the product of BFs for the $\psi(3686)\rightarrow\gamma\chi_{cJ}$,
$\Sigma^{-}\to n\pi^{-}$ and $\bar{\Sigma}^{+}\to\bar{n}\pi^{+}$ channels; and $N_{\psi(3686)}$ is the total number of $\psi(3686)$ events.
The corresponding detection efficiencies and the resultant BFs are summarized in Table~\ref{sum}. We note that due to the low-energy radiative photon of $\chi_{cJ}~(J=1,2)$, the detection efficiency tends to get smaller due to the rejection of $\pi^{0}$-mass requirement.

\begin{table*}[ht]
	\centering
	\footnotesize
	\caption{BFs of $\chi_{cJ}\rightarrow \Sigma^{-}\bar{\Sigma}^{+}$ (in units of $10^{-5}$), where the errors are statistical only.
          The statistical errors of the MC-determined efficiencies are negligible.} 
	\label{sum}
       \begin{tabular}{l C{5cm} C{4cm} R{3cm}}
          \specialrule{1.0pt}{1pt}{1pt}
	\hline
                     \textbf{Quantity}&$\mathbf{\chi_{c0}}$&$\mathbf{\chi_{c1}}$&$\mathbf{\chi_{c2}}$ \\
	\hline	
                     $N^{\rm obs}$ &$2143\pm102$ & $214\pm53$& $131\pm51$\\
    		Efficiency $(\epsilon) \%$ &$9.56$ & $8.58$ & $6.97$\\
    		$\mathcal{B}$$(\psi{(3686)}\rightarrow \gamma \chi_{cJ}) \%$ &$9.79$ & $9.75$ & $9.52$\\
  	          $\mathcal{B}$$(\chi_{cJ}\rightarrow\Sigma^{-}\bar{\Sigma}^{+})(10^{-5})$&$51.3\pm2.4$ &$5.7\pm 1.4$ & $4.4\pm1.7$\\
         	\hline
        \specialrule{1.0pt}{1pt}{1pt}
	\end{tabular}
\end{table*}

\section{\boldmath ESTIMATION OF SYSTEMATIC UNCERTAINTIES}\label{sys1}

Various sources of systematic uncertainties are studied and summarized in Table~\ref{sys}.
The investigated uncertainties are discussed in detail in the following:

\begin{table}[ht]
	\centering
	\footnotesize
	\caption{Systematic uncertainties in the BF measurements in percent.}
	\label{sys}
	\begin{tabular}{l C{2cm} C{1.5cm} C{1.cm}}
	 \specialrule{1.0pt}{1pt}{1pt}
	\hline
		\textbf{Source}&$\mathbf{\chi_{c0}}$&$\mathbf{\chi_{c1}}$&$\mathbf{\chi_{c2}}$ \\
	\hline
                     MDC Tracking& $2.8$& $2.8$& $2.8$\\
                     Photon Reconstruction& $1.0$& $1.0$& $1.0$\\                
                     Kinematic Fit& $5.8$& $5.8$& $5.8$\\
                     $\pi^{0}$ mass window& $1.6$& $-$& $-$\\
                     $\pi^{+}\pi^{-}$ mass window& $0.6$& $-$& $-$\\
                     $M_{\rm rec}({\pi^{+}\pi^{-})}$ mass window& $1.0$& $-$& $-$\\
                     Bin size of $\Sigma^{-}$;& $0.3$& $1.0$& $1.5$\\
                     Signal Shape&$2.6$&$2.8$& $0.0$\\       
	          Background Shape& $1.2$& $2.9$& $3.2$\\
	          Fitting Range&$1.0$&$2.5$&$4.3$\\
	          Signal Shape of $\chi_{cJ}$;&$0.0$& $0.0$& $0.0$\\                    
	          Background Shape&$0.0$&$1.4$&$1.6$\\
	          Fitting Range &$0.2$&$1.8$&$2.3$\\
                     Generator&$-$ & $4.2$&$4.1$\\
                     Truth Match & $0.7$& $0.7$&$0.7$\\
                     Number of $\psi(3686)$&$0.6$&$0.6$&$0.6$\\
                     $\mathcal B(\psi(3686)\to\gamma \chi_{cJ})$& $2.0$& $2.5$& $2.1$\\
                     \hline
                     \addlinespace[0.2ex]
	           Total&$7.9$& $9.8$& $10.2$\\
                     \hline
                       \specialrule{1.0pt}{1pt}{1pt}
	\end{tabular}
\end{table}

~{\it a. MDC Tracking:}~The tracking efficiencies for $\pi^{+}/\pi^{-}$ as functions of the transverse momentum have been studied
with the process $J/\psi\rightarrow \Sigma^{*-}\bar{\Sigma}^{+} \rightarrow\pi^{-}\Lambda~\bar{n}~\pi^{+}(\Lambda\rightarrow\pi^{-}p)$.
The efficiency difference between data and MC is $1.4\%$ for each charged-pion track.

~{\it b. Photon Reconstruction:}~The uncertainty of the photon-detection efficiency is estimated to be $1.0\%$ per photon~\citep{Sys1}. 

~{\it c. $\bar{n}$ Selection and Kinematic Fit:}~The systematic uncertainties of the $\bar{n}$ selection and the kinematic fit involving the $\bar{n}$
is studied using the control sample of $J/\psi\to\Sigma^{*}\bar{\Sigma}^{+}$. The relative difference of 5.8\% in
efficiency between MC and data is assigned as the corresponding systematic uncertainty.

~{\it d. Mass Window Requirement:}~Various cuts in the mass spectra have been used to select events, namely on
$M(\gamma\gamma)$, $M(\pi^{+}\pi^{-})$ and $M_{rec.}(\pi^{+}\pi^{-})$.
Cross checks of systematic effects for these mass window requirements are considered following the procedure described in Ref.~\cite{Sys4}.
The consistency of the results is checked by comparing the uncorrelated differences between the parameter values, $x_{test}\pm\sigma_{test}$,
obtained from the fits to the nominal results, $x_{nom.}\pm \sigma_{nom.}$. The systematic sources cannot be discarded when the significance
of uncorrelated differences, $\Delta x_{uncor.}=|x_{nom.}-x_{test}|/\sqrt{|\sigma^{2}_{nom.}-\sigma^{2}_{test}|}>2$. By comparing the
results of various selections taken within a proper range with the nominal result, the one with the largest difference is taken as an estimate of
the corresponding uncertainty. 
For the $\chi_{c0}$ case, the $\pi^{0}$ veto is tested by varying the rejection windows, $|M(\gamma \gamma)-m(\pi^{0})|$ from 3 to 18$\mevcc$.
The largest deviation $\Delta x_{uncor.}$ is found when the veto is applied at $12\mevcc$.
Similar attempts are performed for the mass windows of $M_{\rm rec}(\pi^{+}\pi^{-})~\rm{and}~M(\pi^{+}\pi^{-})$.
The largest deviations are found when the windows are $| M_{\rm rec}(\pi^{+}\pi^{-}) -m(J/\psi)|>16 \mevcc$
and $| M(\pi^{+}\pi^{-})$ - $m(\textrm{K}^{0}_{\rm S})| >12\mevcc$. The differences to the nominal results are then taken as an estimate of the systematic uncertainty.
In all cases, we observe no tendency of $\Delta x_{uncor.}$ along with the selection variations, indicating no bias in these selection criteria.
For $\chi_{c1,2}$, it is found that the $\Delta x_{uncor.}$ for all the tests are less than $2\sigma$. Therefore, no systematic uncertainties are considered in that case.

~{\it e. Fitting Process:} To estimate the uncertainties from the fitting process, the following three studies are made.

~{\it (i) Bin Width:} The bin width in the bin-by-bin fit is determined to be 10~$\mevcc$ by testing a
series of MC samples as described in Sec.~\ref{extraction}.
The systematic uncertainties are determined by taking the difference between the determined
branching fractions and their input values for $\chi_{cJ}\to\Sigma^{-}\bar{\Sigma}^{+}$.

~{\it (ii). Fit of $\chi_{cJ}$:} To extract the uncertainties associated with the fit procedure on $M_{\rm rec}(\gamma)$, alternative fits are performed
by replacing the second-order polynomial function with a third-order function for the background description, fixing the width of the Gaussian functions
for the signal description, and by varying the fitting range. All the relative changes in the results are taken as the uncertainties from the fit.

~{\it (iii) Fit of $M(\Sigma^{-})$:} Similarly, alternative fits are applied by varying the MC-simulated signal and background shapes and fit ranges.
The differences are treated as a systematic uncertainty.

~{\it f. Generator:} For the $\chi_{c0}$ case, the angular distribution of the $\Sigma^{-}$ in the $\chi_{c0}$ rest frame is
isotropic since the $\chi_{c0}$ is a scalar particle. Therefore, no systematic uncertainty needs to be considered for the $\chi_{c0}$.
For $\chi_{c1,2}$, on the other hand, we considered two extreme cases in the analysis, namely with $\alpha=1$ and $-1$, respectively.
The resulting differences in efficiency with a factor of $\sqrt{12}$ are then assigned as the source of a systematic uncertainty.

~{\it g. MC Truth Matching Angle:} Since in the analysis of the signal MC data sample only events are selected whereby the difference between the angle
of the reconstructed photon and the generated one (MC truth angle) is less than $10^{\circ}$,
it might lead to a systematic error in the efficiency determination.
Several differences with MC truth angles are considered ranging from $10^{\circ}$ to $20^{\circ}$. The largest difference on the efficiencies are
considered as the source of systematic uncertainty.

\begin{table*}[ht]
\footnotesize
\caption{Results of the BFs $(\rm{in~ units~ of ~}10^{-5})$ for the measurement of $\chi_{cJ}\to\Sigma^{-}\bar{\Sigma}^{+}$, compared with
  the $\chi_{cJ}\to\Sigma^{+}\bar{\Sigma}^{-}$ results from BESIII~\citep{BES2} and theoretical predictions~\citep{ppbar}\citep{COM2}\citep{QCM}. The first errors are statistical
  and the second systematic.}
\centering

\begin{tabular}{l C{3cm} C{3cm} C{3cm} C{3cm}C{2cm} c l}
 \specialrule{1.0pt}{1pt}{1pt}
\hline
Channel & This work & Statistical significance & BESIII~\citep{BES2}& \multicolumn{2}{c}{Theoretical predictions}&  \\
\cline{5-6}
 \addlinespace[0.9ex]
            & $\chi_{cJ}\to \Sigma^{-}\bar{\Sigma}^{+}$& &$\chi_{cJ}\to \Sigma^{+}\bar{\Sigma}^{-}$ \
&COM &QCM~\citep{QCM}\\
\hline
$\chi_{c0}\rightarrow \Sigma^{-}\bar{\Sigma}^{+}$ & $51.3\pm2.4\pm4.1$   & $30\,\sigma$ & $50.4\pm2.\
5\pm2.7$ &  5.9-6.9~\citep{COM2}& $18.1\pm 3.9$\\
$\chi_{c1}\rightarrow \Sigma^{-}\bar{\Sigma}^{+}$ &$5.7\pm1.4\pm0.6$ &$5.8\,\sigma$ & $3.7\pm0.6\pm0\
.2$ &  3.3~\citep{ppbar}&...\\%
$\chi_{c2}\rightarrow \Sigma^{-}\bar{\Sigma}^{+}$ &$4.4\pm1.7\pm0.5$ &$3.6\,\sigma$ & $3.5\pm0.7\pm0\
.3$ & $5.0$~\citep{ppbar}&$4.3\pm0.4$\\
\hline
 \specialrule{1.0pt}{1pt}{1pt}
\end{tabular}
\label{result}
\end{table*}

~{\it Other Uncertainties:} The total number of $\psi(3686)$ decays is determined by analyzing the inclusive hadronic events from $\psi(3686)$ decays with
an uncertainty of $0.6\%$~\citep{psidata}. The uncertainties due to the BFs $\psi(3686)\to\gamma \chi_{cJ}$ are quoted from the PDG~\citep{PDG}.
The systematic error due to uncertainties in the trigger efficiency is negligible for this analysis.

~{\it Total Systematic Uncertainty:} We assume that all systematic uncertainties given above are independent and we add them in quadrature to
obtain the total systematic uncertainty.

\section{SUMMARY}
Based on $(448.1\pm 2.9) \times10^{6}\,\psi(3686)$ events collected with the BESIII detector, the BFs of the processes $\chi_{cJ}\rightarrow\Sigma^{-}\bar{\Sigma}^{+}$ are measured and the results are summarized in Table~\ref{result}. This is the first BF measurement of $\chi_{cJ}\rightarrow\Sigma^{-}\bar{\Sigma}^{+}$. The results of $\chi_{cJ}\to \Sigma^{-}\bar{\Sigma}^{+}$ are consistent with $\chi_{cJ}\to\Sigma^{+}\bar{\Sigma}^{-}$~\citep{BES2} from BESIII within the uncertainties, which confirm the prediction of  isospin symmetry. The BF of $\chi_{c0}\to \Sigma^{-}\bar{\Sigma}^{+}$ does not vanish, which demonstrates a strong violation of the HSR. Both predictions based on the COM~\citep{COM2} and QCM~\citep{QCM} fail to describe our measured result. The measured BFs of $\chi_{c1,2}\to\Sigma^{-}\bar{\Sigma}^{+}$ are in good agreement with the theoretical predictions based on COM~\citep{ppbar} and consistent within $1\sigma$ with the prediction based on QCM~\citep{QCM} for $\chi_{c2}\to\Sigma\Sigma$.


\section*{\boldmath ACKNOWLEDGMENTS}
The BESIII collaboration thanks the staff of BEPCII and the IHEP computing center and the supercomputing center of USTC for their strong support. This work is supported in part by National Key Basic Research Program of China under Contract No. 2015CB856700; National Natural Science Foundation of China (NSFC) under Contracts Nos. 11625523, 11635010, 11735014, 11822506, 11835012, 11935015, 11935016, 11935018, 11961141012, 11335008, 11375170, 11475164, 11475169, 11625523, 11605196, 11605198, 11705192; the Chinese Academy of Sciences (CAS) Large-Scale Scientific Facility Program; Joint Large-Scale Scientific Facility Funds of the NSFC and CAS under Contracts Nos.U1732263, U1832207, U1532102, U1832103; CAS Key Research Program of Frontier Sciences under Contracts Nos. QYZDJ-SSW-SLH003, QYZDJ-SSW-SLH040; 100 Talents Program of CAS; INPAC and Shanghai Key Laboratory for Particle Physics and Cosmology; ERC under Contract No. 758462; German Research Foundation DFG under Contracts Nos. Collaborative Research Center CRC 1044, FOR 2359; Istituto Nazionale di Fisica Nucleare, Italy; Ministry of Development of Turkey under Contract No. DPT2006K-120470; National Science and Technology fund; STFC (United Kingdom); The Knut and Alice Wallenberg Foundation (Sweden) under Contract No. 2016.0157; Olle Engkvist Foundation (Sweden); The Royal Society, UK under Contracts Nos. DH140054, DH160214; The Swedish Research Council; U. S. Department of Energy under Contracts Nos. DE-FG02-05ER41374, DE-SC-0012069.



\begin{thebibliography}{99}
\bibitem{explan} Due to the unknown polarization in the production of $\Sigma^{-}$ and small value of asymmetry parameters, $\alpha=0.068$~\citep{PDG}, the efficiency difference between \textsc{PHSP} and differential helicity amplitude can be neglected.
\end{thebibliography}
\end{document}